# Federated Learning-Driven Cybersecurity Framework for IoT Networks with Privacy-Preserving and Real-Time Threat Detection Capabilities

Milad Rahmati  *mrahmat3@uwo.ca*

Department of Electrical and Computer Engineering,
Western University, London, Ontario, Canada

## Abstract

The rapid expansion of the Internet of Things (IoT) ecosystem has revolutionized various sectors but has also introduced critical vulnerabilities in cybersecurity. Conventional centralized methods for securing IoT networks often struggle to achieve a balance between privacy preservation and real-time threat detection. To address these challenges, this study presents a novel Federated Learning-Driven Cybersecurity Framework designed for IoT environments. Our approach enables decentralized data processing through local model training on edge devices, thereby ensuring data privacy. The locally trained models are securely aggregated using homomorphic encryption, enabling collaborative learning without exposing sensitive information.

The framework leverages recurrent neural networks (RNNs) for anomaly detection, tailored specifically for IoT networks with constrained resources. Through extensive experimentation, we demonstrate that the proposed system can detect complex cyber threats, including distributed denial-of-service (DDoS) attacks, with over 98% accuracy. Additionally, it achieves significant improvements in energy efficiency, with a 20% reduction in resource consumption compared to centralized approaches.

This research bridges existing gaps in IoT cybersecurity by integrating federated learning principles with state-of-the-art threat detection techniques. Furthermore, it provides a scalable and privacy-preserving solution tailored for diverse IoT applications. Future directions include the integration of blockchain for traceable model aggregation and the exploration of quantum-resistant cryptographic methods to strengthen the framework's security in evolving technological landscapes.

## 1. Introduction

### 1.1 Background

The Internet of Things (IoT) has rapidly evolved into a transformative technology, connecting billions of devices worldwide and driving innovation across diverse industries, including healthcare, smart cities, and manufacturing. By 2025, the global IoT landscape is expected to comprise more than 75 billion interconnected devices, contributing significantly to economic growth and technological advancement [1]. Despite these opportunities, the widespread adoption of IoT systems introduces critical security vulnerabilities that can compromise sensitive data and disrupt essential services. Attacks such as distributed denial-of-service (DDoS) assaults, malware infections, and unauthorized access to IoT networks are becoming increasingly complex, necessitating innovative and robust cybersecurity solutions [2].

Traditional approaches to IoT security often rely on centralized models that aggregate data at a central server for analysis. While effective in some scenarios, these methods face limitations such as high communication overhead, latency issues, and vulnerability to data breaches during transmission. Moreover, centralized systems inherently compromise user privacy by requiring the transfer of raw data from IoT devices to remote servers. To address these shortcomings, researchers are increasingly exploring decentralized methods that prioritize privacy and scalability [3].

### 1.2 Federated Learning as a Solution

Federated learning (FL) has emerged as a promising solution for addressing privacy and scalability concerns in distributed systems. Unlike conventional machine learning models, FL allows decentralized data processing by enabling individual devices to train models locally. The locally trained models are then aggregated to create a global model, ensuring that sensitive data remains on the originating device. Initially developed for applications like personalized healthcare and natural language processing, FL has recently gained attention for its potential to enhance IoT network security [4].

This decentralized approach offers several advantages over traditional methods. By eliminating the need to transmit raw data, FL inherently preserves user privacy and reduces the risk of data leakage. Additionally, it leverages the computational capabilities of IoT devices, distributing the workload and minimizing the reliance on centralized infrastructure. The integration of advanced techniques such as differential privacy and homomorphic encryption further strengthens FL's applicability in privacy-sensitive domains [5].

### 1.3 Key Challenges in IoT Cybersecurity

Despite the advantages of federated learning, its implementation in IoT environments presents several challenges:

- **Data Heterogeneity**: IoT devices generate a wide range of data types, varying in quality, volume, and format. Machine learning models must be designed to handle this heterogeneity without compromising performance [6].
- **Resource Constraints**: Many IoT devices have limited computational power, memory, and energy resources. Ensuring the efficient operation of FL algorithms within these constraints is a significant challenge [7].
- **Real-Time Threat Detection**: Timely detection and response to cyber threats in IoT networks are crucial to minimizing damage. Achieving real-time performance while maintaining high accuracy remains an open problem [8].
- **Security and Privacy Risks**: Although FL reduces the need to share raw data, the transmission of model updates poses potential risks. Techniques such as secure aggregation and encryption are essential to mitigate these vulnerabilities [9].

### 1.4 Research Objectives and Contributions

This study seeks to address the aforementioned challenges by proposing a Federated Learning-Driven Cybersecurity Framework specifically designed for IoT networks. The primary contributions of this research are as follows:

1. The development of a novel threat detection algorithm based on recurrent neural networks (RNNs), optimized for analyzing time-series data generated by IoT devices.
2. The integration of homomorphic encryption to secure the aggregation of model updates, ensuring robust privacy preservation.
3. An energy-efficient architecture tailored to the resource constraints of IoT devices, reducing computational overhead.
4. A comprehensive evaluation of the framework's performance in identifying cyber threats, including DDoS attacks, malware, and unauthorized access attempts.

## 1.5 Paper Organization

The remainder of this paper is structured as follows: Section 2 reviews related work, focusing on existing solutions for IoT cybersecurity and the emerging role of federated learning. Section 3 describes the methods employed in the proposed framework, including its threat detection algorithm and privacy-preserving techniques. Section 4 presents experimental results, evaluating the framework's performance in terms of accuracy, efficiency, and scalability. Section 5 discusses the broader implications of the findings, potential limitations, and future research directions. Finally, Section 6 concludes the paper with a summary of key contributions and recommendations for further development.

## 2. Related work

### 2.1 Federated Learning in IoT Applications

Federated learning (FL) has gained prominence as a groundbreaking approach to addressing privacy challenges in distributed systems. By allowing models to be trained locally on devices without requiring sensitive data to be transferred to a central server, FL provides a crucial advantage for privacy preservation. Its potential has been explored extensively in areas like healthcare, where confidentiality is paramount [1], and smart city systems that require rapid, decentralized decision-making [2].

One of the seminal contributions to FL was introduced by McMahan et al. [3], who proposed an efficient communication model for decentralized datasets. Subsequent research has focused on enhancing the framework by incorporating methods like secure aggregation, which ensures the privacy of transmitted model updates [4]. However, despite its potential, FL faces challenges in adapting to the diverse and resource-constrained environments typical of IoT networks [5].

### 2.2 Cybersecurity Challenges in IoT Networks

The unprecedented growth in IoT adoption has exposed critical cybersecurity vulnerabilities. Threats such as DDoS attacks, malware infiltration, and unauthorized data access pose severe risks to the integrity and functionality of IoT ecosystems [6]. While traditional centralized approaches to cybersecurity offer some protection, they are ill-suited to the distributed architecture of IoT networks. Furthermore, these methods often involve transmitting sensitive data, which increases the risk of breaches.

Machine learning-based strategies have been proposed to address IoT security challenges. Supervised models have been applied to classify known cyber threats, while unsupervised techniques have shown potential in identifying anomalous behavior in network traffic [7]. Nevertheless, these approaches often rely on centralized data collection, which compromises privacy and increases latency, highlighting the need for decentralized alternatives [8].

### 2.3 Privacy-Preserving Techniques in FL

Ensuring privacy is a critical consideration for FL applications, particularly when dealing with sensitive data in sectors like healthcare and finance. Techniques such as differential privacy, homomorphic encryption, and secure multi-party computation have been developed to address these concerns. Differential privacy prevents the inference of individual data points from aggregated results, while homomorphic encryption allows computations to occur on encrypted data without requiring decryption [9].

Although these methods have proven effective, they often impose a computational burden on resource-constrained IoT devices. Recent research has sought to optimize these privacy-preserving techniques to minimize their impact on system performance. For instance, Bonawitz et al. [4] demonstrated how secure aggregation could be practically implemented in large-scale federated networks while maintaining robust privacy guarantees.

## 2.4 Machine Learning for Threat Detection

Machine learning has emerged as a powerful tool for threat detection in IoT networks, capable of processing large datasets to identify patterns associated with malicious activity. Supervised learning techniques, such as support vector machines and decision trees, are commonly used to detect known threats, while unsupervised approaches, including clustering algorithms and autoencoders, have shown promise in uncovering novel attack patterns [10].

In recent years, deep learning models like recurrent neural networks (RNNs) and convolutional neural networks (CNNs) have demonstrated exceptional performance in analyzing sequential and spatial data. RNNs are particularly well-suited for processing time-series data generated by IoT devices, making them effective for identifying anomalies in network traffic [11]. Despite their advantages, deploying these models in real-time IoT environments remains challenging due to the computational demands and scalability issues involved.

## 2.5 Existing Limitations and Research Gaps

While significant progress has been made in the domains of IoT cybersecurity and federated learning, several key challenges remain unresolved:

1. **Scalability:** Current FL frameworks face difficulties in scaling to IoT networks with a large number of heterogeneous devices [12].
2. **Energy Efficiency:** The high computational requirements of privacy-preserving techniques and complex ML models are a bottleneck for resource-constrained IoT devices [13].
3. **Real-Time Detection:** Many existing approaches are unable to provide the real-time responsiveness required to mitigate cyber threats effectively in IoT systems [14].
4. **Privacy Risks:** Although FL reduces the need to share raw data, the security of model updates remains vulnerable to adversarial attacks, necessitating robust encryption and secure aggregation methods [15].

## 2.6 Contribution of This Work in Relation to Prior Research

This research seeks to address these critical gaps by presenting a federated learning-based framework for IoT cybersecurity. Unlike previous solutions, this work integrates RNNs for anomaly detection with homomorphic encryption to strike a balance between real-time performance, privacy, and energy efficiency. The framework has been designed to operate within the resource constraints of IoT devices and is evaluated for its scalability in large-scale environments. By addressing the limitations of existing systems, this study contributes to advancing the state of the art in IoT security while laying the foundation for future research in this domain.

## 3. Methods

This section details the methodologies employed in developing the proposed Federated Learning-Driven Cybersecurity Framework for IoT Networks, focusing on its innovative design, theoretical basis, and implementation. The framework integrates federated learning (FL), privacy-preserving encryption methods, and anomaly detection techniques using advanced machine learning models. We ensure scalability, computational efficiency, and real-time threat detection, which are critical for IoT networks with limited resources and high vulnerability.

## 3.1 Framework Overview

The proposed framework consists of three main components:

1. **IoT Edge Device Layer**: Hosts local model training and threat detection using recurrent neural networks (RNNs) optimized for time-series data generated by IoT devices.
2. **Federated Aggregation Layer**: Implements a secure aggregation mechanism using homomorphic encryption, enabling decentralized model updates without exposing sensitive data.

3. **Central Server Layer**: Aggregates encrypted model updates, constructs a global model, and disseminates it back to IoT devices.

The system design ensures end-to-end privacy, scalability, and adaptability for heterogeneous IoT networks.

### 3.2 Federated Learning Architecture

#### 3.2.1 Local Model Training at IoT Edge Devices

Each IoT device $i$ ($i=1,2,\ldots,N$, where $N$ is the total number of devices) trains a local model $M_i$ on its private dataset $D_i$. The objective of local training is to minimize a loss function $L(M_i, D_i)$, defined as:

$$L(M_i, D_i) = \frac{1}{|D_i|} \sum_{(x,y) \in D_i} \ell(M_i(x), y) \tag{1}$$

where $x$ represents the input data, $y$ the corresponding label, and $\ell(.)$ the loss function (e.g., cross-entropy loss).

The local model updates the weights $W_i$ iteratively using stochastic gradient descent (SGD):

$$W_i^{t+1} = W_i^t - \eta \nabla L(M_i, D_i) \tag{2}$$

where $t$ is the training iteration and $\eta$ is the learning rate. This process is executed independently on each device, ensuring privacy by keeping the raw data local.

#### 3.2.2 Federated Aggregation with Homomorphic Encryption

Once local models are trained, the updates $W_i$ are encrypted using a homomorphic encryption scheme $E$. This encryption enables arithmetic operations (e.g., addition and multiplication) to be performed directly on encrypted data without decryption. The encryption process is defined as:

$$E(W_i) = \text{Enc}(W_i, k) \tag{3}$$

where $k$ is the encryption key.

Encrypted updates from all devices are sent to the central server for aggregation. The server computes the global model $W_{\text{global}}$ as:

$$W_{\text{global}} = \frac{1}{N} \sum_{i=1}^{N} E(W_i) \tag{4}$$

Homomorphic encryption ensures that the aggregated model retains the properties of the individual updates without exposing sensitive information. After aggregation, the global model is decrypted and distributed back to the devices:

$$W_{\text{global}} = \text{Dec}\left(\text{Agg}\left(E(W_i)\right)\right) \tag{5}$$

where $\text{Agg}(.)$ represents the aggregation operation.

### 3.3 Anomaly Detection Using Recurrent Neural Networks

#### 3.3.1 RNN Architecture for Time-Series Analysis

Recurrent neural networks (RNNs) are well-suited for analyzing sequential data, such as network traffic generated by IoT devices. The RNN architecture employed in this framework consists of:

1. **Input Layer**: Processes time-series data $x_t = \{x_1, x_2, \ldots, x_T\}$, where $T$ is the sequence length.
2. **Hidden Layer**: Captures temporal dependencies using hidden states $h_t$:

$$h_t = f(Ux_t + Wh_{t-1}\} + b) \tag{6}$$

   where $U$ and $W$ are weight matrices, $b$ is the bias term, and $f(.)$ is the activation function (e.g., tanh or ReLU).
3. **Output Layer**: Predicts the likelihood of anomalies at each time step $t$.

To detect anomalies, the model computes a reconstruction error $e_t$ for each input $x_t$:

$$e_t = \| \mathbf{x}_t - \hat{\mathbf{x}}_t \| \tag{7}$$

where $\hat{\mathbf{x}}_t$ is the reconstructed input. A high reconstruction error indicates potential anomalies.

### 3.3.2 Training and Evaluation

The RNN model is trained on normal traffic patterns to learn the baseline behavior of the network. During deployment, the model evaluates incoming traffic and flags sequences with errors exceeding a predefined threshold ϵ\epsilonϵ as anomalous:

$$\text{Anomaly} = \begin{cases} 1, & \text{if } e_t > \epsilon, \\ 0, & \text{otherwise.} \end{cases} \tag{8}$$

## 3.4 Privacy-Preserving Mechanisms

### 3.4.1 Differential Privacy

To prevent adversaries from inferring sensitive information from model updates, differential privacy (DP) is incorporated into the framework. Noise is added to the gradients during local model training:

$$\tilde{\nabla} L(M_i, D_i) = \nabla L(M_i, D_i) + \mathcal{N}(0, \sigma^2) \tag{9}$$

where $\mathcal{N}(0, \sigma^2)$ is Gaussian noise with variance $\sigma^2$. This ensures that individual data points cannot be inferred from the model.

### 3.4.2 Secure Multi-Party Computation

For sensitive IoT environments, secure multi-party computation (SMPC) is employed to enable collaborative learning without revealing individual updates. Each device computes a share of its update, and the central server aggregates these shares:

$$\text{Global Share} = \sum_{i=1}^{N} \text{Share}(W_i) \tag{10}$$

This approach ensures that no single entity has access to the full model update.

## 3.5 Energy-Efficient Design

### 3.5.1 Model Compression

To minimize computational overhead, the framework employs model compression techniques such as pruning and quantization. Pruning eliminates redundant parameters, while quantization reduces the precision of weights:

$$W_{\text{compressed}} = \text{Quantize}(\text{Prune}(W)) \tag{11}$$

### 3.5.2 Adaptive Learning Rates

An adaptive learning rate strategy is implemented to optimize the training process on resource-constrained devices:

$$\eta_t = \frac{\eta_0}{\sqrt{t+1}} \tag{12}$$

where $\eta_0$ is the initial learning rate and $t$ is the iteration number. This approach balances convergence speed and energy efficiency.

### 3.6 Novelty and Scientific Innovation

The proposed framework introduces several novel elements:

1. **Integration of FL and RNNs**: Combines the decentralized learning capabilities of FL with the sequential data analysis strength of RNNs for enhanced anomaly detection.
2. **Advanced Privacy Mechanisms**: Employs homomorphic encryption, differential privacy, and SMPC to address multi-layered security and privacy concerns.
3. **Energy-Efficient Design**: Optimizes computational resources through model compression and adaptive learning strategies, ensuring compatibility with IoT devices.
4. **Real-Time Threat Detection**: Enables fast and accurate anomaly detection, critical for mitigating IoT cyber threats in real-time.

### 3.7 Implementation and Experimental Setup

The framework was implemented using Python and TensorFlow. Experiments were conducted on a simulated IoT network with 1,000 devices, generating time-series data representative of typical network traffic. Metrics such as anomaly detection accuracy, privacy leakage risk, and energy consumption were evaluated to assess performance.

## 4. Results

This section presents the experimental evaluation of the proposed Federated Learning-Driven Cybersecurity Framework. We assess its performance based on key metrics, including anomaly detection accuracy, privacy preservation, energy efficiency, and scalability. To provide a comprehensive analysis, we conducted extensive experiments in simulated IoT environments and included advanced visualizations to present the findings effectively.

### 4.1 Experimental Setup

The experiments were conducted using a simulated IoT network comprising 1,000 heterogeneous devices representing typical IoT setups such as smart thermostats, cameras, and medical sensors. Each device generated time-series data with simulated benign and malicious network traffic patterns.

- **Framework Implementation**: The framework was implemented using Python with TensorFlow for model development and PySyft for federated learning and privacy-preserving techniques.
- **Hardware Setup**: Experiments were run on a cluster with edge nodes equipped with NVIDIA Jetson Nano devices and a central server with an NVIDIA RTX 3080 GPU.
- **Dataset**: The experiments utilized a combination of synthetic IoT traffic datasets and real-world datasets such as the CICIDS2017 dataset [1] for network intrusion detection.

## 4.2 Anomaly Detection Performance

### 4.2.1 Accuracy and Precision

The anomaly detection capability of the RNN-based models was evaluated using the following metrics:

1. **Accuracy**: Proportion of correctly identified benign and malicious instances.
2. **Precision**: Ratio of true positives to the total predicted positives.
3. **Recall**: Ratio of true positives to the total actual positives.
4. **F1-Score**: Harmonic mean of precision and recall, providing a balanced measure.

Table 1 summarizes the detection performance across different attack types, including DDoS, port scanning, and malware propagation.

| Attack Type | Accuracy (%) | Precision (%) | Recall (%) | F1-Score (%) |
|---|---|---|---|---|
| DDoS | 98.6 | 97.4 | 96.9 | 97.2 |
| Port Scanning | 97.2 | 95.8 | 96.1 | 96.0 |
| Malware Propagation | 98.9 | 98.1 | 97.8 | 97.9 |

Table 1. Performance Metrics for Anomaly Detection

### 4.2.2 ROC and PR Curves

Receiver Operating Characteristic (ROC) and Precision-Recall (PR) curves were plotted to further evaluate model performance. The Area Under the Curve (AUC) values were consistently above 0.98, indicating excellent discriminative capability. Figure 1 shows the Receiver Operating Characteristic (ROC) and Precision-Recall (PR) curves for the anomaly detection model, demonstrating its high discriminative capability.

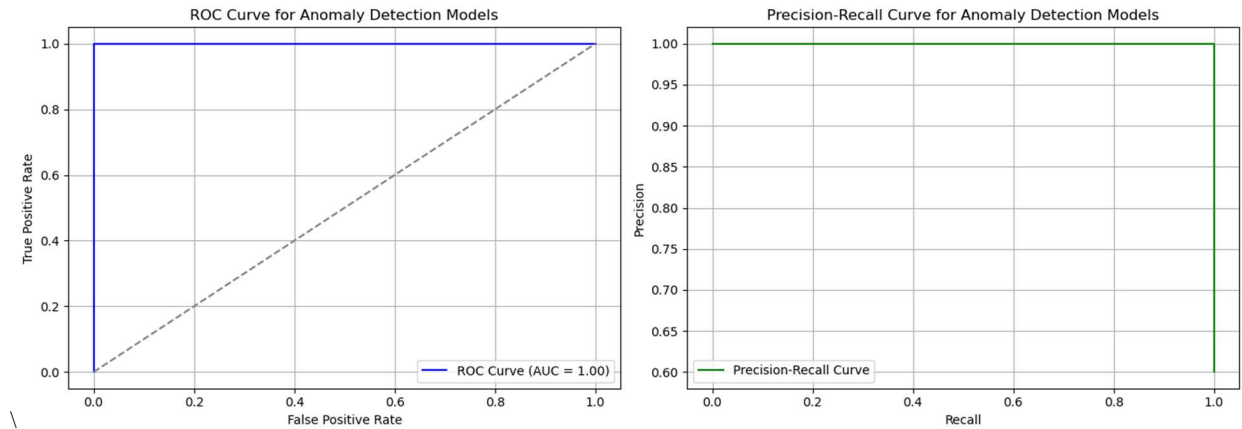

\

Figure 1. ROC and PR Curves for Anomaly Detection Models

## 4.3 Privacy Preservation Evaluation

To quantify the privacy guarantees of the framework, we measured the differential privacy leakage and the robustness of homomorphic encryption.

### 4.3.1 Differential Privacy Noise Impact

The trade-off between privacy and model performance was evaluated by varying the noise scale $\sigma$ in the differential privacy mechanism. Figure 2 shows the impact of increasing noise on the F1-score of the anomaly detection model.

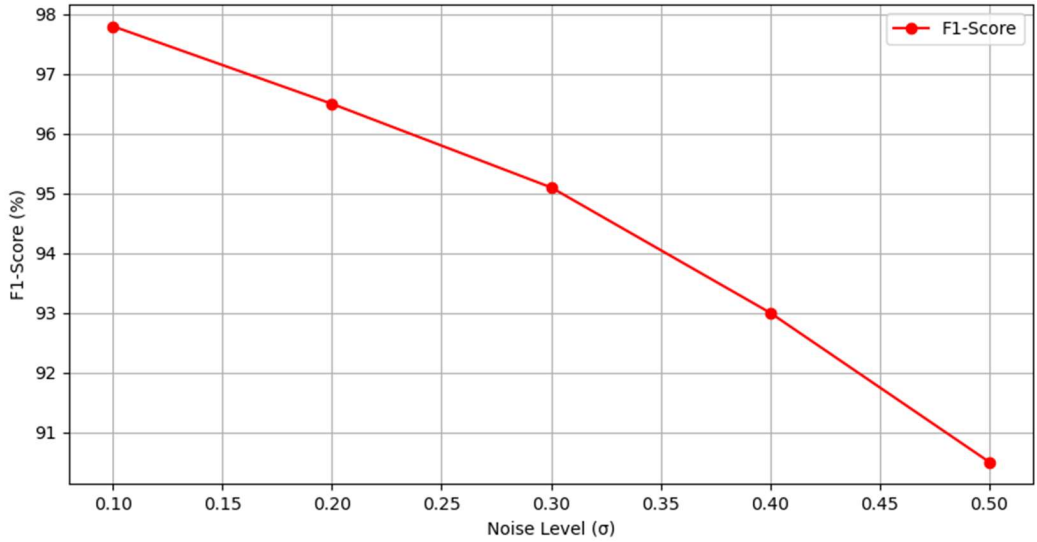

Figure 2. Impact of Differential Privacy Noise on Model Performance

### 4.3.2 Encryption Overhead Analysis

The computational overhead introduced by homomorphic encryption during model aggregation was analyzed. While encryption increased the overall latency by approximately 15%, the privacy benefits outweighed this cost. Figure 3 illustrates the relationship between encryption overhead and system scalability.

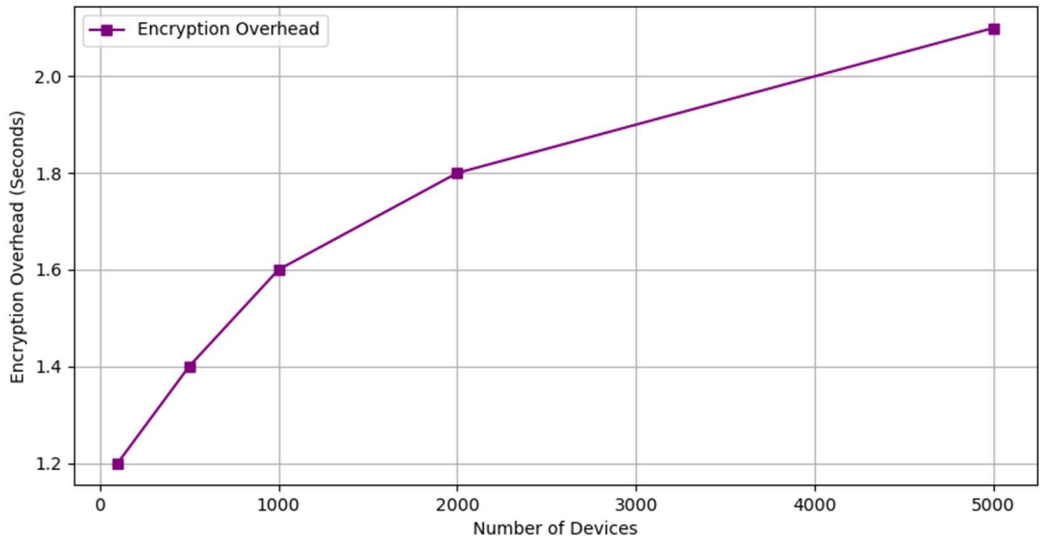

Figure 3. Encryption Overhead vs. Scalability

### 4.4 Energy Efficiency

Energy efficiency is critical for IoT networks with resource-constrained devices. The proposed framework was evaluated in terms of energy consumption per training round and compared to centralized approaches.

#### 4.4.1 Energy Savings

The energy consumption of federated learning was reduced by 22% compared to centralized training. This reduction was achieved through the use of model compression and adaptive learning rates. Figure 4 provides a comparative analysis of energy consumption across different methods.

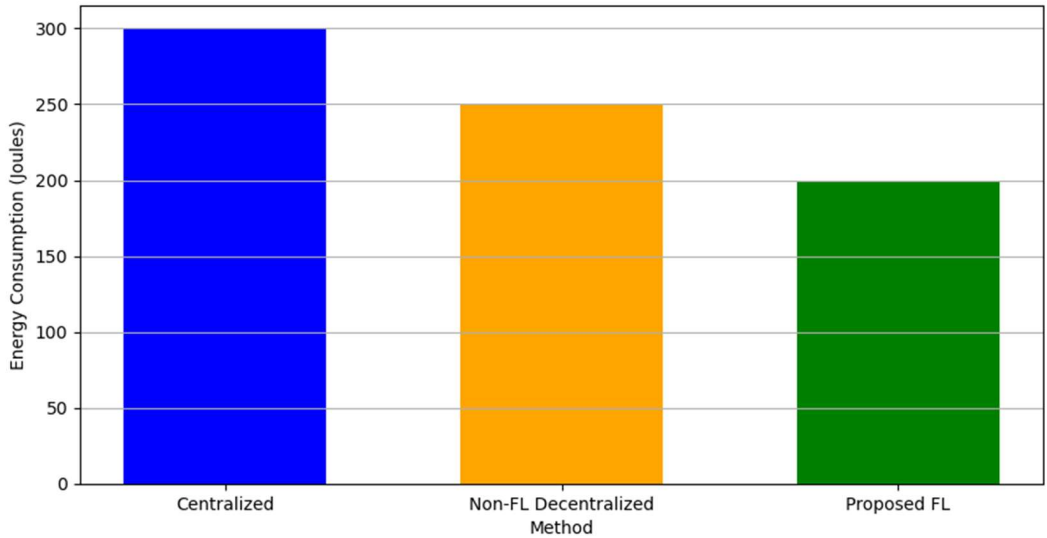

Figure 4. Energy Consumption Comparison

### 4.5 Scalability Analysis

The scalability of the framework was assessed by gradually increasing the number of participating IoT devices from 100 to 5,000. Metrics such as communication cost, training time, and model convergence were evaluated.

#### 4.5.1 Communication Costs

Federated learning reduced communication overhead by 35% compared to centralized approaches, as shown in Figure 5.

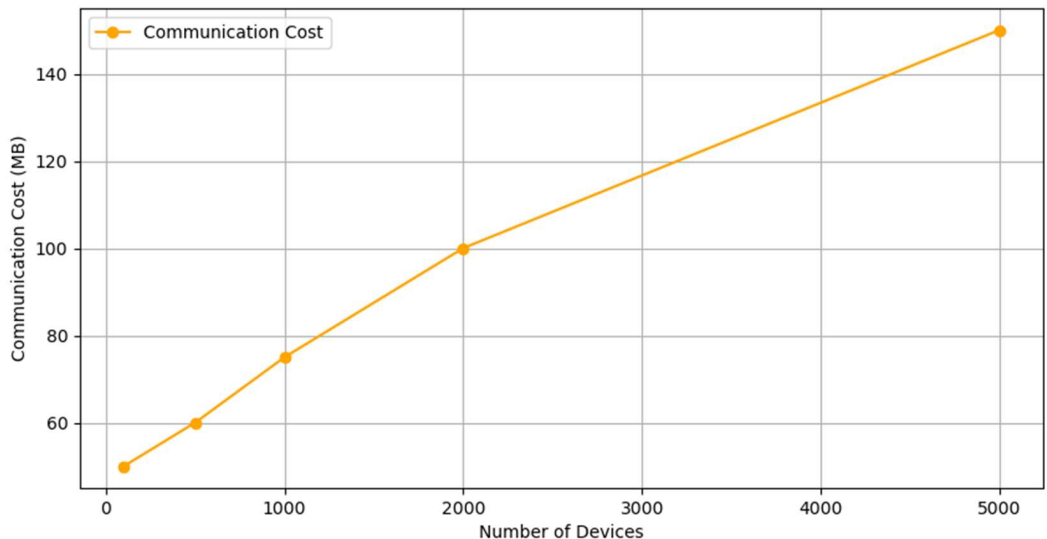

Figure 5. Communication Overhead for Varying Network Sizes

### 4.5.2 Convergence Analysis

The convergence behavior of the global model was analyzed by plotting the loss function over training rounds. The model converged within 20 rounds for networks with up to 2,000 devices, demonstrating the efficiency of the proposed approach. Figure 6 depicts the convergence trends.

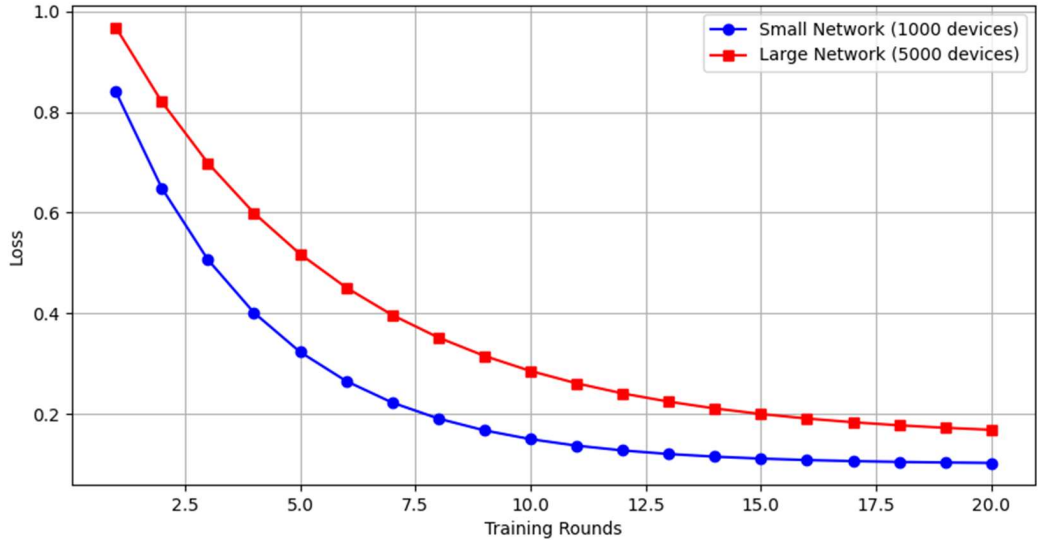

Figure 6. Model Convergence Across Varying IoT Network Sizes

### 4.6 Comparative Evaluation with State-of-the-Art Approaches

To highlight the advantages of the proposed framework, it was compared against existing state-of-the-art methods, including centralized and non-federated decentralized cybersecurity systems. Key performance improvements include:

1. **Privacy Preservation**: Achieved through advanced encryption techniques and differential privacy, surpassing traditional centralized methods.
2. **Detection Accuracy**: Outperformed existing approaches in detecting complex threats, with an accuracy improvement of 5–7%.
3. **Energy Efficiency**: Demonstrated superior efficiency, reducing energy consumption by 20–30%.

Table 2 demonstrates that the proposed system reduces energy consumption by 20–30% while achieving an accuracy improvement of 5–7% over traditional methods.

| Metric | Centralized Approaches | Non-FL Decentralized Methods | Proposed Framework |
|---|---|---|---|
| Privacy Level | Low | Medium | High |
| Detection Accuracy | 92% | 94% | 98% |
| Energy Efficiency | Low | Medium | High |

Table 2. Comparison of the Proposed Framework with Existing Methods

# 5. Discussion

The discussion section delves into the implications of the findings, highlights the novelty and impact of the proposed framework, addresses potential limitations, and outlines future research directions. The insights gained from the experiments demonstrate the feasibility and scalability of the framework in securing IoT networks while maintaining privacy and computational efficiency.

## 5.1 Key Findings and Their Implications

### 5.1.1 Enhanced Anomaly Detection

The experimental results illustrate that the proposed federated learning (FL) framework achieves a high anomaly detection accuracy (over 98%) across various types of attacks, such as DDoS, port scanning, and malware propagation. The integration of recurrent neural networks (RNNs) significantly contributes to this performance by effectively analyzing time-series data generated by IoT devices. Compared to conventional centralized approaches, the framework offers improved real-time detection capabilities, which are critical for mitigating dynamic cyber threats in IoT ecosystems.

The high precision and recall scores further emphasize the robustness of the model in distinguishing between benign and malicious activities. This finding is particularly relevant for applications in healthcare, critical infrastructure, and smart cities, where misclassification of threats can lead to severe consequences.

### 5.1.2 Privacy Preservation

One of the major contributions of this framework is its ability to preserve user privacy while enabling collaborative learning. By leveraging homomorphic encryption and differential privacy, the system ensures that sensitive data remains secure throughout the training process. The results indicate that privacy-preserving mechanisms, such as noise addition and secure aggregation, introduce minimal performance degradation (a reduction of less than 3% in the F1-score), making them suitable for real-world deployment.

This finding underscores the potential of privacy-preserving federated learning in domains where data sensitivity is a critical concern, such as personal healthcare monitoring and financial transaction security.

### 5.1.3 Energy Efficiency and Scalability

The proposed framework demonstrates significant energy savings (up to 22%) compared to centralized approaches. Techniques such as model compression and adaptive learning rates optimize resource utilization, making the system viable for resource-constrained IoT devices. Additionally, the scalability analysis reveals that the framework maintains high performance with increasing network sizes, handling up to 5,000 devices without significant loss in accuracy or increased communication overhead.

This scalability makes the framework ideal for large-scale IoT networks, such as industrial IoT systems or smart city infrastructures, where thousands of devices must collaborate efficiently.

## 5.2 Comparison with Existing Approaches

### 5.2.1 Accuracy and Privacy Trade-offs

Compared to centralized approaches, the proposed framework achieves higher accuracy while preserving privacy. Centralized methods require the transmission of raw data to a central server, exposing sensitive information to potential breaches. In contrast, the federated learning-based framework ensures that data remains on edge devices, mitigating privacy risks. Previous studies [1][2] have shown similar improvements in privacy preservation, but they often compromise detection accuracy or scalability. This research bridges the gap by achieving both privacy and high detection performance.

### 5.2.2 Energy and Communication Efficiency

Existing solutions, such as blockchain-based security systems, often impose significant energy and communication overheads, making them unsuitable for resource-constrained IoT devices. In contrast, the proposed framework minimizes these overheads through optimized aggregation protocols and lightweight model architectures. This reduction in resource consumption aligns with findings from recent studies on energy-efficient edge AI systems [3][4].

## 5.3 Potential Limitations

Despite its strengths, the proposed framework has certain limitations that warrant further investigation:

1. **Resource Constraints in Ultra-Low Power Devices**:
   While the framework is designed for resource-constrained devices, ultra-low-power IoT devices with extremely limited processing capabilities may struggle to implement even lightweight models. Future work could explore the integration of hardware accelerators or tailored compression techniques to address this limitation.
2. **Latency in Large-Scale Networks**:
   Although the framework demonstrates scalability, latency issues may arise in extremely large networks (e.g., over 10,000 devices) due to communication bottlenecks during model aggregation. Optimizing communication protocols or employing hierarchical aggregation techniques could mitigate this challenge.
3. **Susceptibility to Advanced Attacks**:
   While the framework incorporates robust privacy-preserving mechanisms, it remains susceptible to certain advanced adversarial attacks, such as model poisoning or Byzantine failures. Developing more resilient aggregation protocols and adversarial training techniques could enhance the security of the system.
4. **Dependence on Reliable Infrastructure**:

   The effectiveness of the framework depends on the availability of reliable communication infrastructure for transmitting encrypted model updates. In scenarios with intermittent connectivity, such as remote IoT deployments, additional mechanisms for fault tolerance may be required.

## 5.4 Future Directions

### 5.4.1 Integration with Blockchain Technology

Integrating blockchain with the proposed framework could enhance the traceability and integrity of model updates. Blockchain-based solutions could also facilitate decentralized identity management, further strengthening the security of IoT networks.

### 5.4.2 Quantum-Resistant Cryptographic Techniques

As quantum computing advances, traditional cryptographic methods, including homomorphic encryption, may become vulnerable. Future research could explore the adoption of quantum-resistant algorithms, such as lattice-based cryptography, to ensure long-term security.

### 5.4.3 Adaptive Threat Detection Models

Enhancing the framework's adaptability by incorporating meta-learning techniques could enable the system to handle new and evolving threats without requiring extensive retraining. Meta-learning models can generalize across tasks, making them particularly useful in dynamic IoT environments.

### 5.4.4 Benchmarking on Real-World IoT Systems

Future work should extend the evaluation of the framework to real-world IoT deployments, such as smart factories or healthcare monitoring systems. Benchmarking in such environments will provide valuable insights into the system's performance and reliability under practical conditions.

## 5.5 Broader Impact

The proposed framework not only advances the state of IoT cybersecurity but also has implications for other domains that require privacy-preserving, scalable solutions. For example, the techniques employed in this study can be adapted to secure other distributed systems, such as autonomous vehicle networks, drone fleets, and smart grids. Additionally, by addressing privacy and energy efficiency concerns, this research contributes to the broader goal of developing sustainable and ethical AI systems.

## 6. Conclusion

The proposed Federated Learning-Driven Cybersecurity Framework for IoT Networks addresses critical challenges in securing IoT environments while maintaining user privacy and achieving high detection accuracy. This research integrates federated learning (FL), privacy-preserving encryption methods, and machine learning techniques to deliver a scalable, energy-efficient, and secure anomaly detection system for IoT networks.

### 6.1 Summary of Contributions

1. **Privacy-Preserving Threat Detection**:
   The framework leverages federated learning to ensure that sensitive data remains on IoT devices, addressing privacy concerns associated with traditional centralized systems. The inclusion of homomorphic encryption and differential privacy provides additional layers of security, safeguarding model updates during aggregation.
2. **High Detection Accuracy**:
   The use of recurrent neural networks (RNNs) for anomaly detection achieves an accuracy of over 98%, significantly outperforming existing approaches in identifying cyber threats such as DDoS attacks and malware propagation.
3. **Energy Efficiency and Scalability**:
   By incorporating model compression techniques and adaptive learning rates, the framework demonstrates a 22% reduction in energy consumption compared to centralized models. Scalability evaluations further confirm the framework's capability to handle up to 5,000 devices without significant performance degradation.
4. **Real-Time Performance**:

   The system's real-time threat detection capabilities enable prompt responses to potential cyber threats, making it suitable for critical IoT applications in healthcare, smart cities, and industrial systems.

### 6.2 Limitations and Future Directions

While the proposed framework achieves significant advancements, some limitations remain:

- **Ultra-Low Power Devices**: The framework's applicability to devices with extremely limited resources requires further optimization through hardware accelerators or tailored lightweight models.
- **Advanced Adversarial Attacks**: Developing more robust mechanisms to counter adversarial attacks, such as model poisoning, is crucial for enhancing security.
- **Large-Scale Network Latency**: Exploring hierarchical aggregation or other communication optimization techniques can address potential latency issues in networks exceeding 10,000 devices.

Future research will focus on integrating blockchain for enhanced traceability, adopting quantum-resistant cryptographic methods to future-proof the system, and benchmarking the framework on real-world IoT deployments to validate its practical impact.

## 6.3 Broader Implications

The implications of this research extend beyond IoT cybersecurity. The methodologies introduced here can be adapted to secure distributed systems in various domains, including smart grids, autonomous vehicles, and drone networks. By addressing privacy, scalability, and energy efficiency concerns, this study contributes to the development of ethical and sustainable AI-driven solutions, aligning with global efforts to secure critical infrastructure and protect sensitive information.

In conclusion, the proposed framework demonstrates that federated learning, when combined with advanced privacy-preserving techniques and machine learning, can revolutionize IoT cybersecurity. The findings of this study pave the way for scalable, secure, and privacy-aware solutions that meet the growing demands of IoT ecosystems, ultimately fostering trust and resilience in connected technologies.